\documentclass[iop]{emulateapj}

\usepackage{amsmath}
\usepackage{graphicx}
\shorttitle{The GW signal from binary SMBHs}
\shortauthors{Ravi et al.}

\begin{document}

\title{Does a `stochastic' background of gravitational waves exist in the pulsar timing band?}

\author{V. Ravi\altaffilmark{1,2}, J. S. B. Wyithe}
\affil{School of Physics, University of Melbourne, Parkville, VIC 3010, Australia}
\author{G. Hobbs, R. M. Shannon, R. N. Manchester, D. R. B. Yardley, M. J. Keith}
\affil{CSIRO Astronomy and Space Science, Australia Telescope National Facility, PO Box 76, Epping, NSW 1710, Australia}

\altaffiltext{1}{Also at CSIRO Astronomy and Space Science, Australia Telescope National Facility, PO Box 76, Epping, NSW 1710, Australia}
\altaffiltext{2}{E-mail address: v.vikram.ravi@gmail.com}

\begin{abstract}

We investigate the effects of gravitational waves (GWs) from a simulated population of binary super-massive black holes (SMBHs) on 
pulsar timing array datasets. We construct a distribution describing the binary SMBH population from an existing semi-analytic galaxy formation model. 
Using realizations of the binary SMBH population generated from this distribution, we simulate pulsar timing datasets with GW-induced variations. 
We find that the statistics of these variations do not correspond to an isotropic, stochastic GW background. The ``Hellings \& Downs'' correlations 
between simulated datasets for different pulsars are recovered on average, though the scatter 
of the correlation estimates is greater than expected for an isotropic, stochastic GW 
background. These results are attributable to the fact that just a few GW sources dominate the GW-induced variations in every Fourier frequency bin of a 5-year dataset. 
Current constraints on the amplitude of the GW signal from binary SMBHs will be biased. 
Individual binary systems are likely to be detectable in 5-year pulsar timing array datasets where the noise is dominated by GW-induced variations. 
Searches for GWs in pulsar timing array data therefore need to account for the effects of individual sources of GWs.

\end{abstract}


\section{Introduction} 

Detecting and performing science with gravitational waves (GWs) is currently a major goal of experimental astrophysics. Pulsar timing arrays 
\citep[PTAs,][]{hd83,fb90} employ contemporaneous timing observations of millisecond radio pulsars in order to search for the effects of GWs 
perturbing the space-time metric along each pulsar-Earth line of sight. PTAs are complementary to other 
GW detection experiments, such as ground-based and space-based interferometers, in that they are sensitive to GWs in a different frequency band
(currently $\sim5-100$\,nHz, i.e., periods of five years to a few months). In this frequency band, the most promising astrophysical sources of GWs are 
binary super-massive black holes (SMBHs).

SMBHs, with masses in the range $10^{6}M_{\odot}$ to $10^{10}M_{\odot}$,\footnote{This mass range approximately corresponds to the current sample 
of SMBHs with dynamical mass measurements \citep[cf.][]{mmg+11}.} are of great importance to the evolution of galaxies. Feedback from active SMBHs is a 
key element in shaping properties of the observed galaxy population \citep{gbm12}. Models that trace the 
co-evolution of SMBHs and their host galaxies \citep[e.g.,][]{dcs+08} postulate that SMBHs grow primarily through accretion and coalescence with other SMBHs 
during active phases triggered by galaxy mergers. Presumably, there exists a large population of binary SMBHs at various stages of coalescence in the cores of galaxies that have recently merged with other galaxies. The final stages of SMBH-SMBH coalescence are 
driven by losses of energy and angular momentum to GWs, primarily emitted in 
the PTA frequency band. Various works have predicted the average spectrum of the GW strain amplitude from the cosmic population of binary SMBHs 
\citep{jb03,wl03,svc08}. Under the assumption that all binary systems are in circular orbits evolving only through GW emission, this spectrum takes the 
form \citep{p01}
\begin{equation}
h_{c}(f)=A_{1\,\text{yr}}\left(\frac{f}{f_{1\,\text{yr}}}\right)^{-2/3},
\end{equation}
where $h_{c}(f)$ is the characteristic strain as a function of GW frequency, $f$, at the Earth, and 
$A_{1\,\text{yr}}$ is the characteristic strain at a frequency of $f_{1\,\text{yr}}=(1\,\text{year})^{-1}$. The 
characteristic strain is the GW strain per logarithmic frequency interval, and is given by 
\begin{equation}
h_{c}(f)=\sqrt{fS_{h}(f)},
\end{equation}
where $S_{h}(f)$ is the one-sided strain power spectral density (PSD). 

The combined GW signal from binary SMBHs is widely assumed to form an isotropic, stochastic GW 
background. The value of $A_{1\,\text{yr}}$ is used to specify the amplitude of the background. The most recent 
predictions for the value of $A_{1\,\text{yr}}$ were made by \citet{svc08}, who found a likely range of $10^{-16}$ to $2.5\times10^{-15}$. 
This range of predictions was derived by considering a variety of 
SMBH seeding, accretion and feedback scenarios, as well as uncertainties in the galaxy merger rate and in the SMBH mass function. 

PTA projects are based on observations of periodic pulses of radio emission from pulsars using large radio telescopes. These observations 
lead to measurements of pulse times of arrival (ToAs) at the observatories. ToA measurements are typically conducted every few weeks 
over many years. Pulsar timing involves parameters of a physical model for the ToAs being fitted to the measured ToAs, with the differences between the observed 
ToAs and the model predictions being the timing residuals. ToAs can be modeled using combinations of deterministic and stochastic processes. 

GWs incident on the Earth (and on the pulsars) cause shifts in the measured pulse frequencies 
of the pulsars \citep{s78,d79}. For a pulsar, indexed by $p$, with intrinsic rotation frequency $\nu_{p}$, consider a GW-induced shift, 
$\Delta\nu(t,\mathbf{r_{p}})$, to this frequency. This shift is a function both of time, $t$, and the Earth-pulsar direction vector, $\mathbf{r_{p}}$. 
The resulting discrete time-series of GW-induced variations to the ToAs, $\delta^{p}_{i}$ (the $i$ subscript indicates that $\delta^{p}_{i}$ is 
sampled at times $t_{i}$), is given by \citep{d79}
\begin{equation}
\delta^{p}_{i}=\int_{0}^{t_{i}}\frac{\Delta\nu(t',\mathbf{r_{p}})}{\nu_{p}}dt'.
\end{equation}
For any GW signal, the expected values of zero-lag cross-correlations between the $\delta^{p}_{i}$ time-series for different pulsars 
can be specified. For any stochastic GW signal, the expected value of the normalized correlation for each pulsar pair is expressed in terms of the angular 
separations between the pulsars as \citep{hd83,jhl+05}
\begin{equation}
\rho_{pq}=\frac{3}{2}\alpha\log\alpha-\frac{\alpha}{4}+\frac{1}{2} +\frac{1}{2}\delta_{pq},
\end{equation}
where $\alpha=\frac{1}{2}(1-\cos\theta_{pq})$, $\theta_{pq}=\cos^{-1}\frac{\mathbf{r_{p}\cdot r_{q}}}{|\mathbf{r_{p}}||\mathbf{r_{q}}|}$ 
is the angular separation between pulsars $p$ and $q$, and the Kroenecker delta, $\delta_{pq}$, is unity if $p=q$, and zero otherwise. This function is known as the 
Hellings \& Downs curve. An isolated source of GWs will give rise to correlations that are different from the the Hellings \& Downs curve. 
Measurements of correlations between pulsar timing datasets that are attributable to the effects of GWs are necessary for the detection of GWs with 
PTAs \citep{jhl+05,ych+11,dfg+12}.

The prospect of detecting or constraining the amplitude of a background of GWs from binary SMBHs has been the primary rationale for the  
development of the PTA concept. Various works have placed upper bounds on the value of $A_{1\,\text{yr}}$ for a background with the characteristic 
strain spectral form of Equation~1 \citep{jhv+06,vlj+11,dfg+12}. The best published upper bound \citep{vlj+11} finds that $A_{1\,\text{yr}}<6\times10^{-15}$ with 
95\% confidence. All analysis methods developed to study the combined GW signal from binary SMBHs with PTAs assume 
that the $\delta^{p}_{i}$ time-series for multiple pulsars can be described as a specific stochastic process. We describe the exact nature of this assumption in \S2.

In this paper, we elucidate the statistical nature of the ToA variations induced by GWs from binary SMBHs. We accomplish this by 
modeling the GW signal from the predicted population of binary SMBHs, and by simulating realizations of $\delta^{p}_{i}$ corresponding to realizations of the 
GW signal. This study is critical to the validity of interpreting published upper limits on $A_{1\,\text{yr}}$ as representative 
of limits on the mean characteristic strain spectrum of GWs predicted to arise from binary SMBHs. Our results are also important for the 
optimization of GW detection techniques with PTAs. In \S3, we outline 
our method of simulating pulsar timing datasets including GWs from the predicted population of binary SMBHs. 
Our analysis and results are presented in \S4 and \S5, and we discuss the interpretation and implications of our results in \S6. 
We present our conclusions in \S7.

Throughout this work, we assume a $\Lambda$CDM concordance cosmology based on a combined analysis of the first-year WMAP 
data release \citep{svp+03} and the 2dF Galaxy Redshift Survey \citep{cdm+01}, with $\Omega_{M}=0.25$, $\Omega_{b}=0.045$, 
$\Omega_{\Lambda}=0.75$, $\sigma_{8}=0.9$ and $H_{0}= 73$\,km\,s$^{-1}$\,Mpc$^{-1}$. Although these parameter values have since been 
superseded by more recent observations, we adopt them in order to remain consistent with the model we use for the binary SMBH population \citep{gwb+11}. 
A list of important symbols in this paper is shown in Table~1, along with the sections of the text in which they are introduced.

\begin{deluxetable*}{lll}
\tabletypesize{\scriptsize}
\tablecaption{List of symbols.}
\tablewidth{0pt}
\tablehead{
\colhead{Symbol} & \colhead{Section} & \colhead{Description}
}
\startdata
$\delta^{p}_{i}$ & 1 & GW-induced ToA variation for pulsar $p$ at time $t_{i}$. \\
$\rho_{pq}$ & 1 & Expected zero-lag normalized cross-correlation between  $\delta^{p}_{i}$ and $\delta^{q}_{i}$ time-series. \\
$S_{g}(f)$ & 2 & Expected PSD of $\delta^{p}_{i}$ time-series. \\
$\tilde{S^{p}_{k}}$ & 2 & Periodogram estimator of $S_{g}(f)$ at frequency $f_{k}$.\\
$h_{0}$ & 3.1 & GW strain amplitude divided by frequency dependence. \\
$\Phi$ & 3.1 & Binned distribution of binary SMBHs derived from realizations of the Millennium and Millennium-II coalescence lists. \\
$\bar{\Phi}$ & 3.1 & Average of 1000 realizations of $\Phi$. \\
$\Phi_{\text{fit}}$ & 3.1 & Analytic fit to $\bar{\Phi}$. \\
$S_{\text{g},\,\text{fit}}(f)$ & 3.2 & $S_{g}(f)$ derived in terms of $\Phi_{\text{fit}}$. \\
$h_{\text{c},\,\text{fit}}(f)$ & 3.2 & Expected GW characteristic strain spectrum derived in terms of $\Phi_{\text{fit}}$. \\
$W_{i}$ & 4 & 1\,ns rms ToA variation at time $t_{i}$. \\
$D^{p}_{i}$ & 4 & Sum of $W_{i}$ and $\delta^{p}_{i}$. \\
$S(f)$ & 4 & Expected PSD of $D^{p}_{i}$ time-series. \\
$\tilde{\psi^{p}_{k}}$ & 4 & Periodogram estimator of $S(f)$ at frequency $f_{k}$.\\
$\tilde{\rho}_{pq}$ & 5 & Estimator of $\rho_{pq}$. 
\enddata
\end{deluxetable*}

\section{The current model for ToA variations induced by gravitational waves from binary SMBHs}

ToA variations induced by GWs from binary SMBHs ($\delta^{p}_{i}$) are commonly modeled among the PTA community as a wide-sense stationary 
random Gaussian process. This is based on the hypothesis that many GW sources forming a GW background contribute to 
the ToA variations, resulting in a statistical process governed by the central limit theorem. While the nature of the random Gaussian model for 
$\delta^{p}_{i}$ has been extensively described elsewhere \citep[e.g.,][]{vlm+09}, we summarize it here for completeness. 

The key property of a random Gaussian process is that a linear combination of samples will have a joint normal distribution function. Different samples 
need not be statistically independent. The distribution of samples from a (zero-mean) random Gaussian process is characterized by the covariance 
matrix of the samples. Consider a vector, $\mathbf{R_{p}}$, containing $n$ samples of $\delta^{p}_{i}$. That is, 
\begin{equation}
\mathbf{R_{p}}= \left( \begin{array}{c}
\delta^{p}_{0} \\
\delta^{p}_{1} \\
... \\
\delta^{p}_{n-1} \end{array} \right) .
\end{equation}
Let $\mathbf{R_{q}}$ be another vector defined similarly to $\mathbf{R_{p}}$, corresponding to a pulsar $q$, containing $n$ 
simultaneously-obtained samples of $\delta^{q}_{i}$. Under the random Gaussian assumption, 
the joint probability distribution of the samples in $\mathbf{R_{p}}$ and $\mathbf{R_{q}}$, 
which we denote $P_{pq}$, is given by
\begin{equation}
P_{pq}=\frac{1}{\sqrt{(2\pi)^{n}\text{det}(\mathbf{C_{pq}}})}e^{-\frac{1}{2}\mathbf{R_{p}}^{T}\mathbf{C_{pq}}\mathbf{R_{q}}}.
\end{equation}
Here, $\mathbf{C_{pq}}$ is an $n\times n$ matrix containing the covariances between the samples of $\delta^{p}_{i}$ and $\delta^{q}_{i}$; that is, 
element $ij$ of $\mathbf{C_{pq}}$ is given by the covariance between $\delta^{p}_{i}$ and $\delta^{q}_{j}$. As the Gaussian process is wide-sense stationary, 
each element $ij$ of $C_{pq}$ depends only on the time difference $\tau_{ij} = |t_{i}   - t_{j}|$ between samples $i$ and $j$ for pulsar $p$ and $q$ respectively. Elements 
of $\mathbf{C_{pq}}$ are sampled from a covariance function, $c_{pq}(\tau)$, between the GW-induced ToA variations for pulsars $p$ and $q$. This covariance 
function is defined by the inverse Fourier transform of the one-sided PSD, $S_{g}(f)$, of GW-induced ToA variations for a given pulsar:
\begin{equation}
c_{pq}(\tau)=\rho_{pq}\text{Real}[\mathcal{F}^{-1}(S_{g}(f))].
\end{equation}
Here, $\mathcal{F}$ denotes a complex Fourier transform, and $\tau$ is a time-lag. The PSDs of the GW-induced ToA variations for all pulsars are equivalent, and 
given by \citep{jhv+06}
\begin{equation}
S_{g}(f)=\frac{1}{12\pi^{2}}\frac{h_{c}^{2}(f)}{f^{3}},
\end{equation}
for a GW signal with the expected characteristic strain spectrum $h_{c}(f)$.

The above discussion applies equivalently if pulsar $p$ and pulsar $q$ are the same pulsar, or if they are different pulsars. The Hellings \& Downs 
factor $\rho_{pq}$, defined in Equation~4, accounts for the correlation between GW-induced TOA variations for different pulsars. 
If $h_{c}(f)$ has the form in Equation~1, we have $S_{g}(f)\propto f^{-13/3}$. The GW-induced ToA variations for 
each pulsar will therefore be a ``red'' process. In this work, we only consider time-series $\delta^{p}_{i}$ with finite lengths $T_{\text{obs}}$.

We are interested in comparing a new model for $\delta^{p}_{i}$ with the random Gaussian model described in this section. To this end, we 
need to be able to simulate realizations of $\delta^{p}_{i}$ as a random Gaussian process. Multiple PTA groups test their data analysis algorithms by 
simulating realizations of $\delta^{p}_{i}$ using the GWbkgrd plugin \citep{hjl+09} to the {\sc tempo2} pulsar timing package \citep{hem06}.
While this plugin does not explicitly generate random Gaussian realizations of $\delta^{p}_{i}$ by construction, we and others \citep{vlj+11,dfg+12} 
have checked that it approximates a random Gaussian process well.  

In the plugin GWbkgrd, a number of GW oscillators, $N_{T2}$, are 
simulated between GW frequencies $f_{\text{lo}}$ and $f_{\text{hi}}$, with the normally distributed $+$ and $\times$ GW polarization amplitudes set to be purely real 
with zero mean, variance  
\begin{equation}
\sigma_{T2}^{2}=\sqrt{\frac{\ln(f_{\text{hi}}/f_{\text{lo}})}{N_{T2}}}h_{c}(f),
\end{equation}
and frequency probability distribution, $dP/df$, given by 
\begin{equation}
\frac{dP}{df}=\frac{1}{\ln(f_{\text{hi}}/f_{\text{lo}})}f^{-1}.
\end{equation}
ToA variations calculated for a given pulsar $p$ at different times $t_{i}$ corresponding to GWs from each of these oscillators are summed to produce 
a realization of the $\delta^{p}_{i}$ time-series. The frequency limits $f_{\text{lo}}$ and $f_{\text{hi}}$ are generally chosen respectively to be much less than the 
$T_{\text{obs}}^{-1}$ and much greater than the Nyquist frequency corresponding to the minimum sampling interval. 

We make a distinction between the expected PSD of ToA variations induced by GWs from binary SMBHs, as defined in Equation~8, 
and estimates of this PSD based on realizations of the ToA variations. 
A commonly-used non-parametric, unbiased estimator of the PSD of a time-series is the periodogram \citep{s898}. The periodogram, $\tilde{S^{p}_{g}}$, 
of $\delta^{p}_{i}$ is defined as 
\begin{equation}
\tilde{S^{p}_{k}}=\frac{2}{T_{\text{obs}}}|\text{DFT}[\delta^{p}_{i}]|^{2},
\end{equation}
where $\text{DFT}$ denotes a discrete Fourier transform. We adopt the following standard definition for the $\text{DFT}$ of $n$ samples of $\delta^{p}_{i}$:
\begin{equation}
\text{DFT}(f_{k})=\sum_{m=0}^{n-1}\delta^{p}_{m}e^{-i2\pi mk/n}\frac{T_{\text{obs}}}{n},
\end{equation}
where $i=\sqrt{-1}$ in this case. The DFT is evaluated for frequencies 
\begin{equation}
f_{k}=(k+1)\frac{1}{T_{\text{obs}}},\,\,0\leq k<\frac{T_{\text{obs}}}{2T_{\text{samp}}}, 
\end{equation}
where $T_{\text{samp}}$ is the interval (assumed to be constant) between samples of $\delta^{p}_{i}$.
Throughout this work, we estimate the PSD, $S_{g}(f)$, of realizations of $\delta^{p}_{i}$ by evaluating $\tilde{S^{p}_{k}}$.

\section{Simulating pulsar ToA variations accounting for binary SMBH population characteristics} 

In this section, we describe a new method of simulating ToA variations caused by GWs from the predicted population 
of binary SMBHs. Various works have presented 
models for the cosmic demographics of binary SMBHs \citep{dsd12}. More recently, such efforts have been based on analytic 
prescriptions for baryon physics applied to dark matter halo merger tree catalogues from N-body simulations \citep[][hereafter G11]{gwb+11}. 
In particular, the G11 prescriptions were applied to merger trees from both the Millennium \citep{swj+05} and the 
Millennium-II \citep{bsw+09} simulations. The Millennium and Millennium-II simulations follow the evolution of dark matter 
structures, using the same physical prescriptions and number of particles. The Millennium-II simulation was however carried out in a 
comoving cubic volume with one-fifth the side length as that of the Millennium simulation, with the aim of resolving smaller-scale dark matter structures than the 
Millennium simulation.\footnote{The Millennium-II simulation however does not reproduce larger-scale structures as well as the Millennium simulation.} 
Together, these simulations resolve dark matter halos corresponding to the observed galaxy population, 
from dwarf galaxies to the largest-mass early-type galaxies. 

The G11 model is the latest in a series \citep{swj+05,csw+06,db07} of semi-analytic prescriptions applied to 
the Millennium simulations. A host of observables of galaxies at low redshifts are reproduced, along with the redshift-evolution of the quasar population 
and star formation. Of most relevance here is that the model also traces the SMBH population, reproducing the $z=0$ SMBH-galaxy 
scaling relations in their slopes, normalization and scatters, as well as the inferred SMBH mass function \citep{mbb+08}. 

We base our description of the binary SMBH population emitting GWs on the prediction for the SMBH-SMBH coalescence rate from the G11 model. 
We fit an analytic function to the distribution of binary SMBHs, and randomly draw GW sources from this distribution to produce realizations of the 
GW sky corresponding to binary SMBHs. We then add the effect of each GW source to simulated pulsar ToA datasets in order to analyze the 
GW-induced ToA variations.

This work is different from previous attempts to model the GW signal from binary SMBHs. Initial attempts \citep[e.g.,][]{jb03} to predict the 
mean GW characteristic strain spectrum from binary SMBHs used empirical determinations of the galaxy merger rate and the SMBH mass function. 
\citet{wl03} predicted the GW spectrum by analytically following the dark matter halo merger hierarchy in an 
extended Press-Schechter framework, and by deriving the SMBH coalescence rate by relating the SMBH masses to the halo 
circular velocities. \citet{svc08} considered the possible range of predictions of the characteristic strain spectrum, using Monte Carlo 
realizations of dark matter halo merger trees and various prescriptions for SMBH growth. 

The key difference between the present work and previous calculations of the 
GW signal from binary SMBHs is that we are chiefly concerned 
with the statistics of $\delta^{p}_{i}$. Our approach to modeling the binary SMBH population is similar to \citet{svv09} 
in our use of mock galaxy catalogues derived from analytic prescriptions applied to the Millennium simulations. However, whereas 
\citet{svv09} modeled the SMBH population by using empirical SMBH-galaxy scaling relations combined with (earlier) mock catalogues, we utilize SMBHs 
modeled by G11 in a self-consistent framework which reproduces the relevant observables. 

\subsection{Modeling the distribution of binary SMBHs}

As in the previous works discussed above, we consider all binary SMBHs to be in circular orbits, and use expressions for the resulting GW emission presented 
by \citet{t87}. We briefly discuss the assumption of circular orbits in \S6.3. The strain amplitude, $h_{s}(f)$, at frequency $f$ of GWs from a circular binary, averaged 
over all orientations and polarizations, is given by:
\begin{equation}   
h_{s}(f)=\sqrt{\frac{128}{15}}\frac{(GM_{C})^{5/3}}{c^{4}D(z)}(\pi f(1+z))^{2/3},
\end{equation}
where $G$ is the universal gravitational constant, $M_{C}=(M_{1}M_{2})^{3/5}(M_{1}+M_{2})^{-1/5}$ is the 
chirp mass of the binary, $c$ is the vacuum speed of light, $z$ is the redshift, and $D(z)$ is the comoving distance 
to the binary. The evolution of the received GW frequency with observed time, $t$, is determined by 
\begin{equation}
\frac{df}{dt}=\frac{96}{5}c^{-5}\pi^{8/3}f^{11/3}(GM_{C}(1+z))^{5/3}. 
\end{equation}
The rest-frame binary orbital frequency is given by $f_{b}=\frac{1}{2}f(1+z)$. 
We assume, after \citet{h02}, that the maximum received frequency, $f_{\text{max}}$, is attained at a binary separation of 
three Schwarzschild radii, corresponding to the last stable orbit:
\begin{equation}
f_{\text{max}}\approx\frac{c^{3}}{12\sqrt{3}\pi(1+z)GM_{C}},
\end{equation}
assuming a mass ratio of unity.

The mock galaxy catalogues resulting from the G11 model are available online\footnote{http://www.mpa-garching.mpg.de/millennium/}  \citep{lv06}.
The halo merger trees from the Millennium and Millennium-II simulations were evaluated at 60 logarithmically-spaced redshift ``snapshots'' between 
$z=0$ and $z=20$. We obtained the lists of SMBH-SMBH coalescence events within the comoving volume of each simulation by querying the online 
database. Redshifts at the (non-logarithmic) midpoints between the redshift snapshots were assigned to each event. We used these lists to fill bins of a 
distribution, $\Phi$, of the number, $N$, of observable binary SMBHs per unit comoving volume per solid angle on the sky, given by
\begin{equation}
\Phi=\frac{dN}{dh_{0}}4\pi\frac{d^{2}V_{c}}{d\Omega dz}\frac{dz}{dt}\frac{dt}{df}
\end{equation}
where 
\begin{equation}
h_{0}=\left(\frac{(GM_{C})^{5/3}}{c^{4}D(z)}(\pi (1+z))^{2/3}\right)^{2}=(\sqrt{\frac{15}{128}}h_{s}f^{-2/3})^{2},
\end{equation}
and $4\pi\frac{d^{2}V_{c}}{d\Omega dz}$ is the sky-integrated comoving volume shell between redshifts $z$ and $dz$. 
Also, $\frac{dz}{dt}=H_{0}\sqrt{\Omega_{M}(1+z)^{3}+\Omega_{\Lambda}}$, and the derivative $\frac{dt}{df}$ was obtained from 
Equation~15. $\Phi$ is the predicted distribution of binary SMBHs in $h_{0}$ (which corresponds to the frequency-independent 
GW ``power'', or squared strain amplitude) and the observed GW frequency. 

For chirp masses below $10^{7}M_{\odot}$, the limited capability of the Millennium simulation to resolve low-mass halos caused an under-prediction of the chirp mass function as 
compared to the Millennium-II simulation. In order to ensure a complete chirp mass function, we included  
binary SMBHs with $M_{C}>10^{7}M_{\odot}$ from the Millennium list of coalescence events, and binaries with 
$10^{6}M_{\odot}<M_{C}\leq10^{7}M_{\odot}$ from the Millennium-II list. 

Some degrees of randomization in the coalescence lists were possible. First, in cases where more than two SMBHs coalesced 
to form a single SMBH between redshift snapshots, the merger order was not specified. In these instances we randomized 
over merger order. Second, a spherical comoving volume shell between any pair of redshifts less 
than $\sim0.09$ could be contained within the simulation volume. Some Millennium redshift snapshots exist at $z<0.09$, and the comoving 
volume shells between redshifts corresponding to these snapshots enclose some SMBH-SMBH coalescence events in the G11 model. 
An observer located at the center of the Millennium simulation volume would observe only a fraction of the total list of events in the G11 model 
at $z<0.09$, and an observer located elsewhere in the volume would observe a different selection of events. This is not the case for the Millennium-II 
simulation, where the volume was too small to enclose any comoving volume shells between redshift snapshots. 
For each realization of the Millennium (but not the Millennium-II) coalescence list, we therefore specified randomly-placed 
spherical shells within the simulation box to select binary SMBHs at these redshifts. For coalescence events at $0.09<z<0.19$, the corresponding comoving volume shells 
between redshift snapshots were smaller than the Millennium simulation volume, though not enclosed by it. For these coalescence events, we randomly included binaries 
in the Millennium coalescence list according to probabilities given by the ratios between the volumes of the comoving shells and the Millennium simulation volume. 
We used 1000 realizations of the Millennium and Millennium-II coalescence lists to form realizations of the binary SMBH distribution $\Phi$. 

In generating realizations of the distribution $\Phi$, we assumed that every SMBH-SMBH coalescence event in the G11 model 
catalogues was the result of a binary SMBH system that had decayed through GW emission. The G11 model included the assumption 
that, upon the merger of two galaxies with central SMBHs, the SMBHs coalesced in every case, before accretion onto the newly-formed 
SMBH.\footnote{There are various mechanisms by which extreme mass ratio binary SMBH systems and triple or higher-order systems can avoid 
coalescence \citep[e.g.,][]{vhm03}.} We note that SMBHs with masses as low as 
$10^{3}M_{\odot}$ were present in the G11 model catalogues, but were not included in the $\Phi$ distributions. We verified that relaxing the lower 
cutoff on the SMBH masses in the $\Phi$ distributions from $10^{6}M_{\odot}$ to $10^{3}M_{\odot}$ did not significantly modify the total signal.

The 1000 realizations of $\Phi$ were averaged to form a distribution $\bar{\Phi}$. We fitted $\bar{\Phi}$ with an analytic function which could be used 
to generate random realizations of the observable binary SMBH population. We did not use realizations of $\Phi$ as realizations of the 
binary SMBH population because the $\Phi$-distributions were binned for computational purposes. A four-parameter function,
\begin{equation}
\Phi_{\text{fit}}=n\left(\frac{h_{0}}{p_{h}}\right)^{\alpha}\left(1+\frac{h_{0}}{p_{h}}\right)^{\beta}f^{-11/3},
\end{equation}
with free parameters $n$, $p_{h}$, $\alpha$ and $\beta$, was found to fit $\bar{\Phi}$ well. 
We performed the fit on the logarithm of the data to approximate linearity in the fitting procedure. The best-fit parameters are given in Table~2. 
The frequency-exponent was held fixed at $-11/3$, as predicted by Equations 15 and 17.

\subsection{Realizations of pulsar ToAs with GW-induced variations}

For a set of binary SMBHs drawn from the distribution $\Phi_{\text{fit}}$, we simulated a corresponding time-series $\delta^{p}_{i}$ by summing 
the contributions from each individual binary. Details of the method used to calculate these contributions 
are presented in \citet{hjl+09}. For each binary, we randomized over the right ascension and declination, the orbital inclination angle, the orientation 
of the line of nodes, and the orbital phase angle at the line of nodes. A new publicly-available {\sc tempo2} plugin, ``addAllSMBHBs'', was written to perform this simulation. 

For most of the present work, we did not use {\sc tempo2} to fit timing model parameters. 
Instead, we made use of the tools available for spectral analysis of timing residuals. In our simulations, the ``timing residuals'' corresponded exactly to $\delta^{p}_{i}$ 
given the absence of timing model fitting. 

We consider it important to emphasize the distinction between the $\delta^{p}_{i}$ time-series and the timing residuals resulting from analyses of 
observed ToA datasets. 
Consider a set of observed ToAs that exactly match a particular timing model, except for the addition of GW-induced variations (a $\delta^{p}_{i}$ time-series). Given that 
an observer does not actually possess any prior knowledge of the timing model parameters, the observer will fit the model parameters to the ToAs. The 
resulting timing residuals will not be equivalent to $\delta^{p}_{i}$. This is because the $\delta^{p}_{i}$ variations in the ToAs can alter the apparent 
pulsar timing parameters. For example, the presence of a $\delta^{p}_{i}$ time-series consisting of a sinusoidal signal with a period of one year 
will alter the apparent pulsar position. 

In order to investigate the statistics of $\delta^{p}_{i}$ given our model for the binary SMBH population, we first used {\sc tempo2} to generate 
500 ToAs spanning 5 years exactly corresponding to the PSR J0437$-$4715 timing model \citep{mhb+12}. We then added realizations 
of the $\delta^{p}_{i}$ time-series evaluated at the observed ToAs to these datasets, that is, with $T_{\text{obs}}=5$\,yr and $T_{\text{samp}}=0.01$\,yr. 
The pulsar distance was set to 1\,kpc, and the position was held fixed for all simulations. As the binary SMBHs used to produce realizations of $\delta^{p}_{i}$ had 
randomized positions and orientations, allowing the pulsar position to vary between realizations would not alter our results. The results presented in this paper 
are not dependent on the timing model used or on the pulsar distance from the Earth. We also added Gaussian white noise 
variations with 1\,ns rms to the ToAs. This white noise component is much smaller than is usually observed in ToA datasets, but was necessary to 
smooth over machine precision errors. 

\begin{deluxetable}{ll}
\tabletypesize{\scriptsize}
\tablecaption{Best-fit parameter values of $\Phi_{\text{fit}}$.}
\tablewidth{0pt}
\tablehead{
\colhead{Parameter} & \colhead{Value}
}
\startdata
$n$ & $2087\pm365$ \\
$p_{h}$ & 4.878$\times10^{-23}$$\pm4.45\times10^{-24}$ \\
$\alpha$ & $-$1.72249$\pm0.00064$ \\
$\beta$ & $-$0.3473$\pm0.0046$ 
\enddata
\end{deluxetable}

We included GW sources between $10^{-9}$\,Hz and $10^{-6}$\,Hz in our simulations of $\delta^{p}_{i}$. The lower frequency cutoff was chosen to be less than one fifth 
of $f_{0}=(5\,\text{years})^{-1}$. The upper frequency cutoff was chosen to be greater than $f_{249}=(0.02\,\text{years})^{-1}$. We assumed, after previous works, that 
all GW sources between these frequency cutoffs are non-evolving over a 5-year timespan, i.e., they do not evolve in frequency by more than 
(5\,years)$^{-1}$. The upper bound on the $h_{0}$-values of sources, 
$h_{0,\,\text{max}}(f)$, was set by the last stable orbit of binary SMBHs. We identified this bound by fitting a power law to the high-$h_{0}$ edge of 
the $\bar{\Phi}$ distribution. The lower bound on $h_{0}$, $h_{0,\,\text{min}}$, was set by the lowest non-zero $h_{0}$-value in $\bar{\Phi}$. This value corresponds 
to a binary SMBH containing two $10^{6}M_{\odot}$ components at $z\approx6$. The distribution included more than $6.5\times10^{18}$ GW sources within this 
$h_{0}-f$ domain; the vast computational cost involved makes it impossible to simulate $\delta^{p}_{i}$ with this many sources. Fortunately, the shape of the 
$\Phi_{\text{fit}}$ distribution was such that, at a given frequency, the highest-$h_{0}$ sources 
contributed most to $S_{g}(f)$ (we return to this point below), defined in terms of $\Phi_{\text{fit}}$ (using Equation~8) as:
\begin{equation}
S_{\text{g},\,\text{fit}}(f)=\frac{1}{12\pi^{2}f^{2}}\int_{h_{0,\,\text{min}}}^{h_{0,\,\text{max}}(f)}\Phi_{\text{fit}}h_{s}^{2}(f)dh_{0},
\end{equation}
The average characteristic strain spectrum derived from $\Phi_{\text{fit}}$ is 
\begin{equation}
h_{\text{c},\,\text{fit}}(f)=\left(f\int_{h_{0,\,\text{min}}}^{h_{0,\,\text{max}}(f)}\Phi_{\text{fit}}h_{s}^{2}(f)dh_{0}\right)^{1/2}.
\end{equation}
We found a function, $\hat{h}_{0}(f)$, such that 
\begin{equation}
0.9S_{\text{g},\,\text{fit}}(f)=\int_{\hat{h}_{0}(f)}^{h_{0,\,\text{max}}(f)}\frac{\Phi_{\text{fit}}h_{s}^{2}(f)}{12\pi^{2}f^{2}}dh_{0}=\hat{S}_{\text{g},\,\text{fit}}(f).
\end{equation}
Thus, the GW sources in the domain $\hat{h}_{0}(f)<h_{0}<h_{0,\,\text{max}}(f)$ contribute, on average, 90\% of $S_{g}(f)$ at every frequency. 
Between $10^{-9}$\,Hz and $10^{-6}$\,Hz, this amounted to $\sim4.5\times10^{6}$ sources. We refer 
to this $h_{0}-f$ domain as the ``90\% domain''. The 90\% domain, along with $h_{0,\,\text{max}}(f)$, $\hat{h}_{0}(f)$ and $h_{0,\,\text{min}}$, is 
shown in Figure~1.

\begin{figure}
\centering
\includegraphics[angle=-90, scale=0.75]{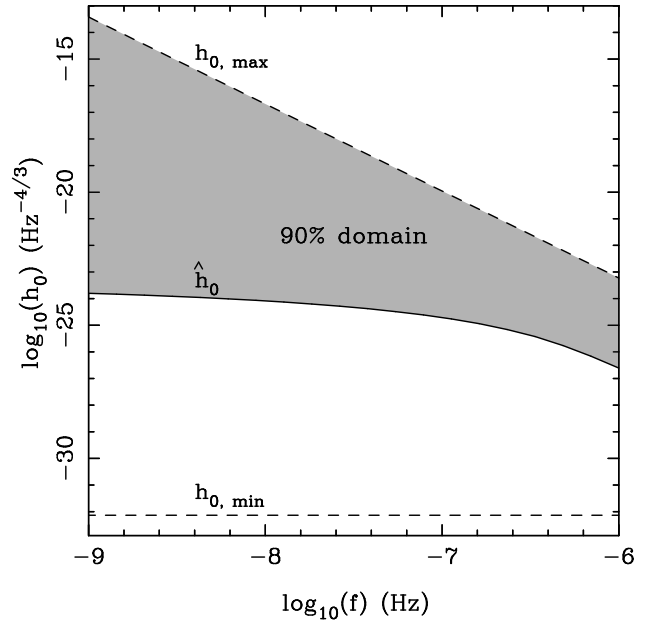}
\caption{Illustration of the $h_{0}-f$ domain constraints on the distribution $\Phi_{\text{fit}}$. The upper and lower dashed lines represent $h_{0,\,\text{max}}(f)$ and 
$h_{0,\,\text{min}}$, as labelled, and the solid curve represents $\hat{h_{0}}(f)$. The shaded region is the ``90\% domain'' from which binary SMBHs 
contributing, on average, 90\% of the ToA variation PSD at every frequency were drawn.}
\end{figure}

We approximated the total number of sources ($6.5\times10^{18}$) in the $h_{0}-f$ domain between $h_{o,\,min}$ and $h_{0,\,\text{max}}$ and 
between $10^{-9}$\,Hz and $10^{-6}$\,Hz as constant. For a given realization of $\delta^{p}_{i}$, the actual number of sources in the 90\% domain is 
governed by binomial statistics. We therefore drew a (binomial-)random number of sources from the 90\% domain, and added contributions from each of them to each realization of $\delta^{p}_{i}$. We assumed that the sources remaining in the $\Phi_{\text{fit}}$ distribution with $h_{0,\,\text{min}}\leq h_{0}\leq \hat{h}_{0}(f)$, 
contributing on average 10\% to $S_{g}(f)$ at every frequency, resulted in a stochastic contribution to $\delta^{p}_{i}$ governed by the central limit theorem. 
We therefore simulated them as described in \S2 using the method of simulating ToA variations corresponding to a GW background implemented in the 
{\sc tempo2} plugin GWbkgrd.  We simulated $N_{T2}=5\times10^{4}$ sources between $10^{-9}$\,Hz and $10^{-6}$\,Hz using the {\sc tempo2} 
method, with the characteristic strain spectrum given by $h_{c}(f)=\sqrt{12\pi^{2}f^{3}(S_{\text{g},\,\text{fit}}(f)-\hat{S}_{\text{g},\,\text{fit}}(f))}$. For each 
realization of $\delta^{p}_{i}$, we added contributions from the  $\sim4.5\times10^{6}$ GW sources drawn from the 90\% domain of the $\Phi_{\text{fit}}$ 
distribution, and from the $5\times10^{4}$ GW sources corresponding to the remaining (on average) 10\% of $S_{g}(f)$ drawn using the {\sc tempo2} method. 
Shifts in the measured pulse frequencies caused by metric perturbations at both the Earth and the pulsar (i.e., the ``Earth term'' and the 
``pulsar term'') were included in our simulations. 

\begin{figure}
\centering
\includegraphics[angle=-90,scale=0.75]{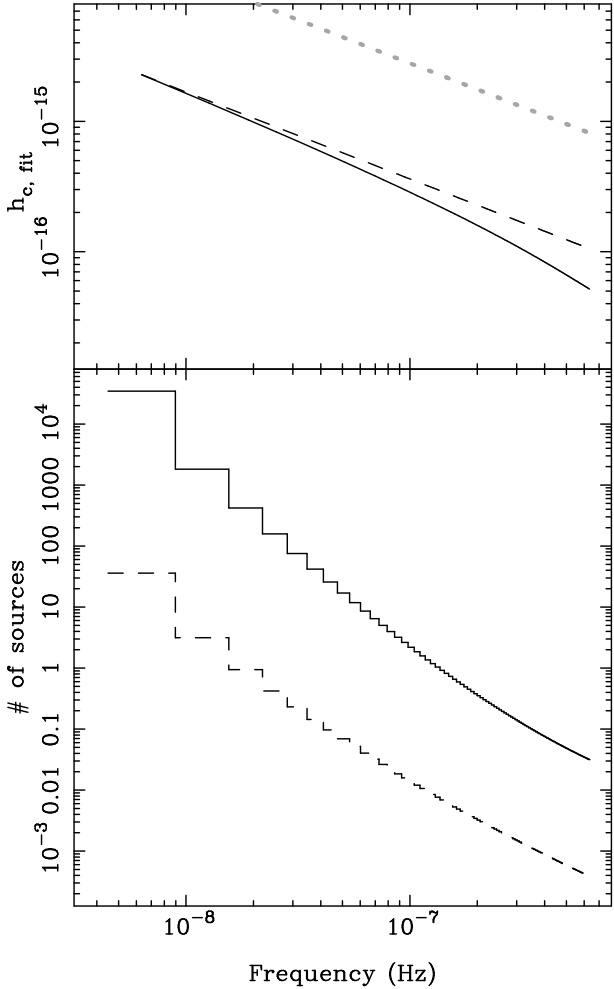}
\caption{\textit{Top:} The solid curve shows the mean characteristic strain spectrum, $h_{\text{c},\,\text{fit}}$, derived from the distribution $\Phi_{\text{fit}}$ in Equation~21. The 
dashed line shows a representative spectrum of the form in Equation~1, with $A_{1\,\text{yr}}=7.8\times10^{-16}$. 
The values of both traces are equivalent at the lowest frequency. 
The dotted line shows a spectrum of the form in Equation~1 with $A_{1\,\text{yr}}=6\times10^{-15}$, corresponding to the most recently published 
95\% confidence upper bound on $A_{1\,\text{yr}}$ \citep{vlj+11}.  
\textit{Bottom:} The numbers of GW sources that contribute 50\% (dashed line) and 90\% (solid line) of $S_{\text{g},\,\text{fit}}(f)$. The numbers are integrated 
over frequency bins of width (5 years)$^{-1}$\,Hz.}
\end{figure}

The top panel of Figure~2 shows $h_{\text{c},\,\text{fit}}(f)$ in the $0\leq k <100$ spectral bins. We also show a characteristic strain spectrum of the form in 
Equation~1 with $A_{1\,\text{yr}}=7.8\times10^{-16}$. This value of $A_{1\,\text{yr}}$ can be taken to be the prediction from the G11 model, as it corresponds 
to the characteristic strain in the lowest ($k=0$) spectral bin of a 5-year dataset. 
This particular curvature in the $h_{\text{c},\,\text{fit}}$ curve, also predicted by \citet{wl03}, is caused by the 
frequency-dependence of $h_{0,\,\text{max}}(f)$, which represents the bound beyond which binary SMBHs have 
crossed the last stable orbit. This curvature is caused by different processes from the curvature reported by \citet{svc08}. We show the mean characteristic 
strain spectrum for GWs from all binary SMBHs in the predicted distribution $\Phi_{\text{fit}}$. \citet{svc08} fitted a broken power-law to realizations of the characteristic 
strain spectrum, accounting for various randomizations over the source population. In particular, \citet{svc08} randomized over the existence of 
``fractional'' sources in every frequency bin of a fiducial dataset, and also excluded the strongest single source in every frequency bin in an attempt 
to isolate the background signal. The smaller number of sources per unit frequency at higher GW frequencies, combined with the greater 
contributions to the signal from the strongest single sources in frequency bins at higher frequencies, both resulted in the curved characteristic 
strain spectra presented by \citet{svc08}.

The bottom panel of Figure~2 shows the mean numbers of highest-$h_{0}$ sources that contribute 90\% and 50\% of $S_{\text{g},\,\text{fit}}(f)$ in these frequency bins. 
A small number of sources contribute a large fraction of $S_{\text{g},\,\text{fit}}(f)$ at every frequency. In the 
$k=0$ frequency bin, the $\sim3\times10^{4}$ highest-$h_{0}$ sources contribute on average 90\% of $S_{\text{g},\,\text{fit}}(f)$, and only 30 
sources on average contribute 50\% of $S_{\text{g},\,\text{fit}}(f)$. At frequencies $f>1.5\times10^{-7}$\,Hz, the strongest source, on average, 
contributes more than 90\% of $S_{\text{g},\,\text{fit}}(f)$ in each frequency bin. This is a consequence of the shallow power-law nature of the 
$h_{0}$-distribution of the GW sources in the $\Phi_{\text{fit}}$ distribution.

In this work, we compare the Millennium-based simulations of $\delta^{p}_{i}$ with simulations of $\delta^{p}_{i}$ created using the 
{\sc tempo2} method described in \S2. To this end, we simulated ToAs as before, but added realizations of $\delta^{p}_{i}$ corresponding to 
$5\times10^{4}$ oscillators simulated using the {\sc tempo2} plugin GWbkgrd. These oscillators were simulated as described in \S2 with a mean characteristic 
strain spectrum given by $h_{\text{c},\,\text{fit}}(f)$. We refer to this latter method of simulating $\delta^{p}_{i}$ as Case H09, after \citet{hjl+09}. Simulations of $\delta^{p}_{i}$ 
using GW sources drawn from $\Phi_{\text{fit}}$ will be referred to as Case R12 after the present work.

\section{Fourier-spectral analysis, and results}

In this section, we consider the differences between the cases in the distributions of the periodograms, $\tilde{S^{p}_{k}}$, evaluated for realizations of $\delta^{p}_{i}$ 
for a single pulsar. This is motivated by the results in Figure~2, in particular that the number of GW sources per spectral bin 
that contribute 90\% of $S_{\text{g},\,\text{fit}}(f)$ varies significantly with frequency.
The Case R12 simulations are intended to more accurately represent the effects of GWs from binary SMBHs on ToA datasets 
than the Case H09 simulations. Example realizations of 5-year $\delta^{p}_{i}$ time-series in both cases are shown in Figure~3. The time-series 
appear quite similar: realizations in both cases are dominated by low-frequency components. Values of up to 1\,$\mu$s are also present in one realization.

\begin{figure}
\centering
\includegraphics[angle=-90, scale=0.75]{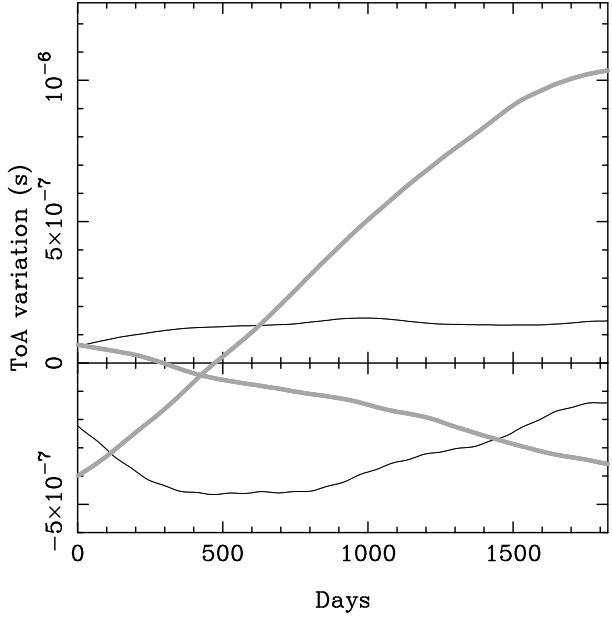}
\caption{Example realizations of $\delta^{p}_{i}$ in Case H09 (thick grey lines) and Case R12 (thin black lines).}
\end{figure}

Instead of directly measuring $\tilde{S^{p}_{k}}$, the added white noise component in the simulated ToAs 
required us to analyze the periodograms of a time-series, $D^{p}_{i}$, given by
\begin{equation}
D^{p}_{i}=\delta^{p}_{i}+W_{i},
\end{equation}
where $W_{i}$ is a time-series of Gaussian white 1\,ns rms ToA variations as discussed above. The PSD of $\delta^{p}_{i}$, $S_{\text{g},\,\text{fit}}(f)$, 
is significantly red, with a spectral index of $-13/3$ (see Equations 1, 20), and is expected to dominate the PSD of $W_{i}$ at low frequencies. 
We used the generalized least-squares algorithm described in \citet{chc+11} to measure the periodograms, $\tilde{\psi^{p}_{k}}$ of realizations of $D^{p}_{i}$. 
This method requires knowledge of the auto-covariance function of $D^{p}_{i}$, which we obtained as in Equation~7 using the inverse DFT of the known PSD 
of $D^{p}_{i}$, $S(f)$, given by 
\begin{equation}
 S(f) = S_{\text{g},\,\text{fit}}(f) + \frac{2(1\,\text{ns})^{2}}{250/(5\,\text{years})}.
 \end{equation}
In the following, we consider the distributions of $\tilde{\psi^{p}_{k}}$ in the lower spectral bins, where $S(f) \approx S_{\text{g},\,\text{fit}}(f)$, to be approximately 
equivalent to the distributions of $\tilde{S^{p}_{k}}$.


\begin{figure}
\centering
\includegraphics[angle=-90,scale=0.75]{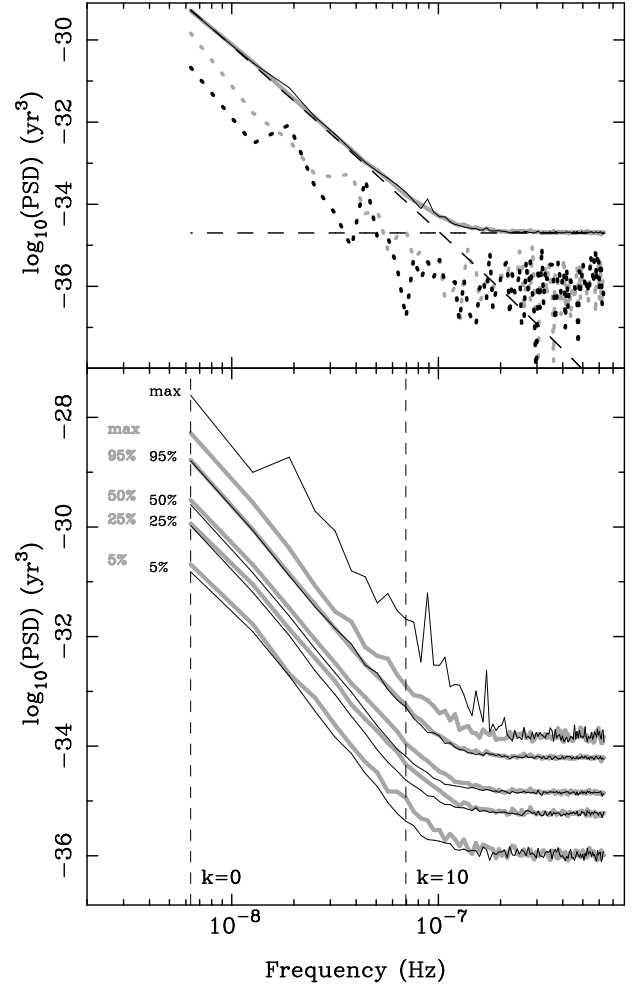}
\caption{\textit{Top:} The mean estimates ($\tilde{\psi^{p}_{k}}$) of the PSD of the simulated ToA variations ($D^{p}_{i}$) in Case R12 (thin solid black line) and in 
Case H09 (thick solid grey line). The predicted PSDs of $\delta^{p}_{i}$ ($S_{\text{g},\,\text{fit}}(f)$) and $W_{i}$ are shown as sloped and horizontal dashed lines respectively. 
Randomly-chosen single measurements of $\tilde{\psi^{p}_{k}}$ in Case R12 and Case H09 are also shown, scaled down by a factor of 10, as black and grey 
dotted lines respectively. \textit{Bottom: } The thin black and thick grey curves depict ``percentile periodograms'' of the distributions of Case R12 and 
Case H09 measurements of $\tilde{\psi^{p}_{k}}$ respectively. The 5th, 25th, 50th, 95th percentiles are shown as labelled, along with the maximum 
values of the periodograms in each spectral bin (labelled ``max''). 
The vertical dashed lines indicate the $k=0$ and $k=10$ spectral bins, with frequencies given by $(k+1)$(5 years)$^{-1}$\,Hz.}
\end{figure}

We produced 1000 realizations of $D^{p}_{i}$ in Case R12 and in Case H09, and measured $\tilde{\psi^{p}_{k}}$ for each realization. 
In the top panel of Figure~4, we show the averages of $\tilde{\psi^{p}_{k}}$ measured 
from each of the Case R12 and Case H09 realizations, along with the expected PSDs of $\delta^{p}_{i}$ and $W_{i}$. Arbitrarily chosen single realizations 
of $\tilde{\psi^{p}_{k}}$ in each case are also shown. The means of the periodograms in both cases are clearly the same, and equivalent to the predicted PSD, $S(f)$, 
given in Equation~24. Though this is as expected, it is both a check of the simulations of $D^{p}_{i}$, and a demonstration of the ability of our PSD estimation method 
to measure steep red spectra without bias.

In contrast, the distributions of $\tilde{\psi^{p}_{k}}$ in the frequency bins where $S(f) \approx S_{\text{g},\,\text{fit}}(f)$ are different between the cases. 
The single realizations of $\tilde{\psi^{p}_{k}}$ in each case shown in the top panel of Figure~4 
begin to hint at these differences. In most spectral bins, the Case R12 periodogram is 
below the Case H09 periodogram. That this is a genuine trend is confirmed in the bottom panel of Figure~4. Here, we depict various ``percentile'' periodograms 
of the distributions of $\tilde{\psi^{p}_{k}}$ in each of Case R12 and Case H09. The percentile periodograms may be interpreted as contours of equivalent 
percentiles of the periodogram distributions in different spectral bins. For example, the ``50\%'' percentile periodogram links the 50th percentile points of the 
distributions of periodogram values in each spectral bin. Below the 95th percentile, all Case R12 percentile periodograms lie below Case H09 percentile periodograms. 
This implies that in most spectral bins, most measurements of a periodogram in Case R12 will, like the individual ones shown 
in the top panel of Figure~4, be below most Case H09 periodograms. However, the 95th percentile periodograms are essentially 
equivalent, and the maximum value Case R12 periodogram is well above the maximum value Case H09 periodogram. These implied ``tails'' at high values 
in the Case R12 periodogram distributions in each spectral bin are highlighted in Figure~5, which depicts the distributions of the $\tilde{\psi^{p}_{k}}$ in 
the spectral bins indicated by the vertical lines in the bottom panel of Figure~4. The distributions are shown as the fractions of Case R12 and Case H09 
periodograms at or above a given value. 

Figure~5 also shows that the Case R12 periodogram distribution in the $k=10$ spectral bin has a longer tail relative to the 
Case H09 distribution, as compared to the $k=0$ spectral bin. This effect is also evident in the bottom panel of 
Figure~4, in that the fractional differences between the percentile periodograms are greater at the upper end of the GW-dominated frequency regime. 
This is consistent with the number of GW sources per spectral bin included in the Case R12 simulations going down with increasing frequency, as shown in Figure~2.

In summary, approximating $\tilde{\psi^{p}_{k}}$ with $\tilde{S^{p}_{k}}$ as discussed above, we find that:
\begin{itemize}
\item In most spectral bins, most realizations of $\tilde{S^{p}_{k}}$ in Case R12 will be below most realizations of $\tilde{S^{p}_{k}}$ in Case H09.
\item The maximum possible values of $\tilde{S^{p}_{k}}$ in Case R12 will be higher than the maximum possible values of $\tilde{S^{p}_{k}}$ in Case H09.
\end{itemize}

\begin{figure}
\centering
\includegraphics[angle=-90,scale=0.75]{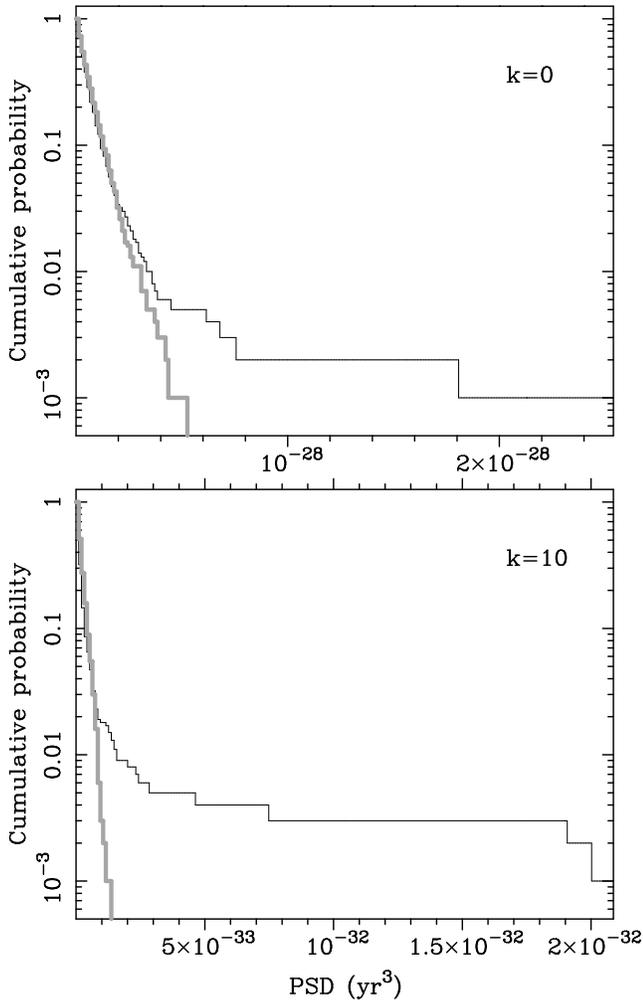}
\caption{The distributions 1000 measurements of $\tilde{\psi^{p}_{k}}$ in Case R12 (thin black lines) and in Case H09 (thick grey lines) in the $k=0$ (top) and $k=10$ (bottom) 
spectral bins. The distributions are shown as the fractions of realizations at or above a given value.  The domains of both plots indicate the maxima and minima of the distributions.}
\end{figure}

\section{Estimates of correlations between GW-induced ToA variation time-series}

\begin{figure*}
\centering
\includegraphics[angle=-90]{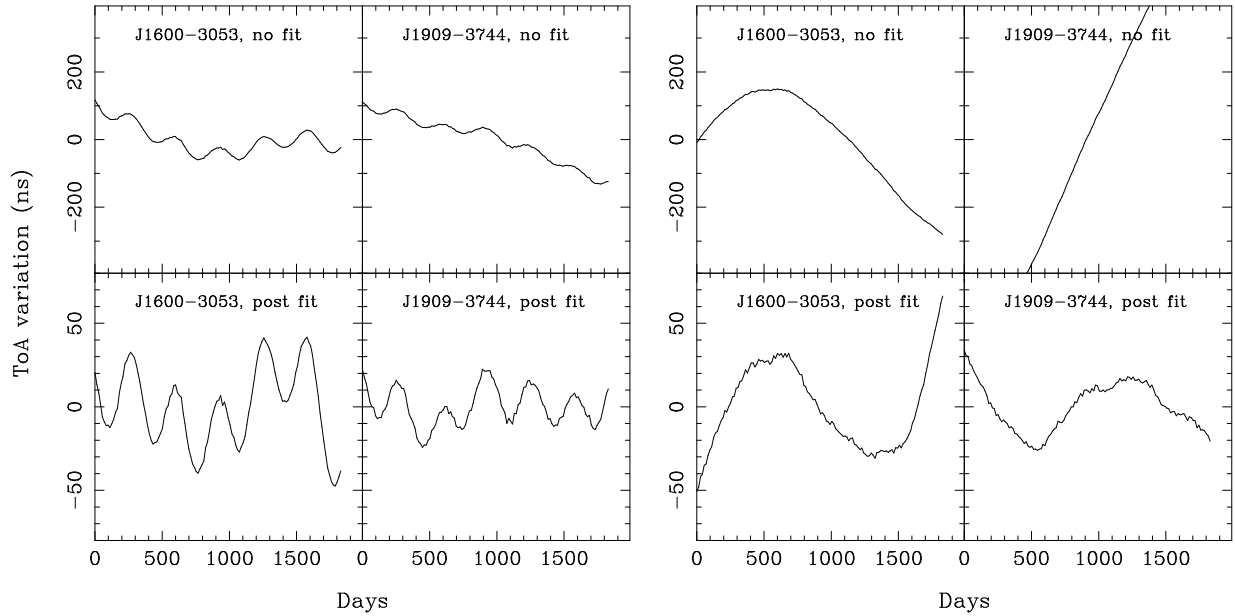}
\caption{Two examples of simulated realizations of $D^{p}_{i}$ for two pulsars: PSR J1600$-$3053 and PSR J1909$-$3744 (see text for details). 
The realization in the left panel is affected by a strong individual GW source, whereas the realizations in the right panel is not. The lower plots 
show the $D^{p}_{i}$ time-series from the corresponding upper plots with linear and quadratic terms removed.}
\end{figure*}

\begin{figure*}
\centering
\includegraphics[angle=-90,scale=0.8]{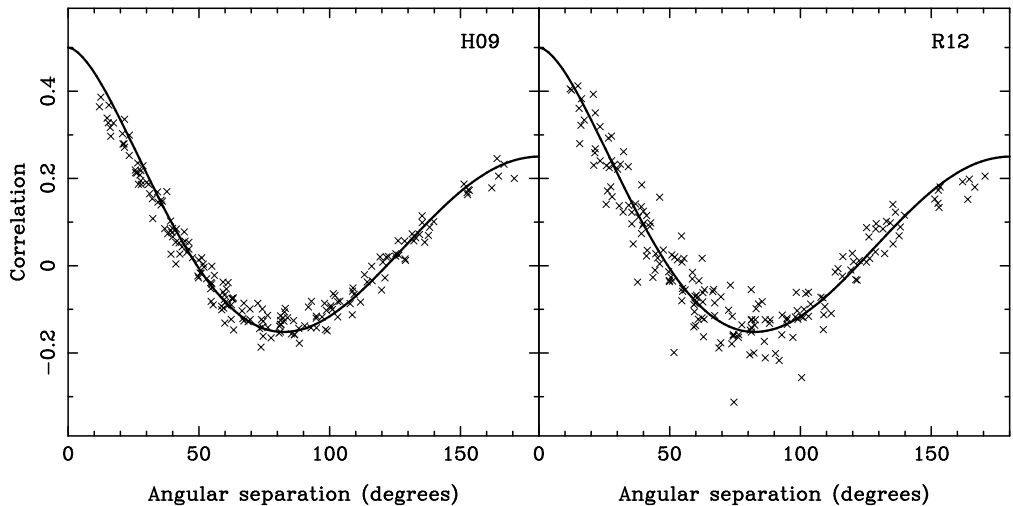}
\caption{The estimated correlations between GW-induced ToA variations for simulated pulsars at the positions of the 20 Parkes PTA pulsars 
in Case H09 (left) and in Case R12 (right), plotted against the angular separations on the sky between each pair of pulsars. 
Each point represents an average over 99 realizations; in Case R12, one realization including an extremely strong individual source was not included in the average. 
Linear and quadratic terms were removed from each ToA variation time-series. The solid curve is the expected Hellings \& Downs curve given in Equation~4. As 
no autocorrelations were present, the maximum value of the Hellings \& Downs curve is 0.5.}
\end{figure*}

\citet{hd83} showed that the \textit{average} values of correlations between $\delta^{p}_{i}$ for different pulsars, for a stochastic GW signal, 
will always be given by the Hellings \& Downs curve (Equation~4). The distributions of individual estimates of these correlations, 
much like the distributions of the PSD estimator $\tilde{S^{p}_{k}}$ considered above, will however depend on the nature of the GW signal. 
Here, we characterize the distributions of estimates of these correlations for multiple pulsar pairs in each case discussed above. 
We simulated 100 realizations of each of Case H09 and Case R12 ToAs as described in \S3.2 for pulsars at 
the positions of each of the 20 pulsars timed by the Parkes PTA, using the timing models specific to each pulsar \citep{mhb+12}. For each realization, 
the same set of GW sources was used to simulate GW-induced ToA variations for each pulsar. The pulsar distances were set arbitrarily between 1\,kpc and 20\,kpc.

We estimated the correlations between time-series $\delta^{p}_{i}$ and $\delta^{q}_{i}$, $\rho_{pq}$, for each pulsar pair $pq$ in each realization of Case R12 and 
Case H09 ToAs. No autocorrelations were estimated. A frequency-domain estimation technique, based on the method of \citet{ych+11}, was used. 
This technique will be detailed elsewhere (Hobbs et al., in preparation). We refer to our estimates of $\rho_{pq}$ as $\tilde{\rho}_{pq}$. 
These estimates were performed using $D^{p}_{i}$ time-series, rather than $\delta^{p}_{i}$ time-series (see Equation~23), and the estimation 
technique was optimized using the expected PSD of $D^{p}_{i}$ given in  
Equation~24. The technique removes best-fit linear and quadratic terms from each $D^{p}_{i}$ time-series 
using the standard {\sc tempo2} least-squares fitting algorithm. This mimics the effect of fitting pulse frequency and frequency-derivative terms to the 
simulated ToAs and then analyzing the timing residuals. 

Following the removal of linear and quadratic terms from one of the 100 Case R12 realizations of $D^{p}_{i}$, a single GW source was found to 
dominate the residual time-series. We show the corresponding $D^{p}_{i}$ time-series for two of the 20 simulated pulsars for this realization in Figure~6. 
The left panel of this Figure shows the large sinusoidal oscillations induced by the source, and the right panel shows example Case R12 realizations of $D^{p}_{i}$ 
that are not dominated by an individual source. It is possible that an individual GW source with a period greater than the 5-year dataspan could dominate the 
realizations of $D^{p}_{i}$ in the right-hand panel of Figure~6. The ToA variations induced by such a source would however be absorbed in the removal of 
the linear and quadratic terms from the $D^{p}_{i}$ time-series.

We averaged all measurements of $\tilde{\rho}_{pq}$ for each pulsar pair $pq$ from the Case R12 realizations, besides the one clearly dominated by an individual source. 
The realization dominated by an individual source added a large amount of scatter to the average Case R12 correlations, and was left out of the average 
to enable a better comparison between the cases.  
We also averaged the Case H09 measurements of $\tilde{\rho}_{pq}$ for each pair $pq$ in 99 arbitrarily-chosen realizations. 
The average measurements of $\tilde{\rho}_{pq}$ are shown for both cases in Figure~7. 
The functional form of the Hellings \& Downs curve is recovered in both Case R12 and Case H09. However, the Case R12 estimates are 
significantly more scattered about the expected values of the correlations than the Case H09 estimates.

\begin{figure}
\centering
\includegraphics[angle=-90,scale=0.7]{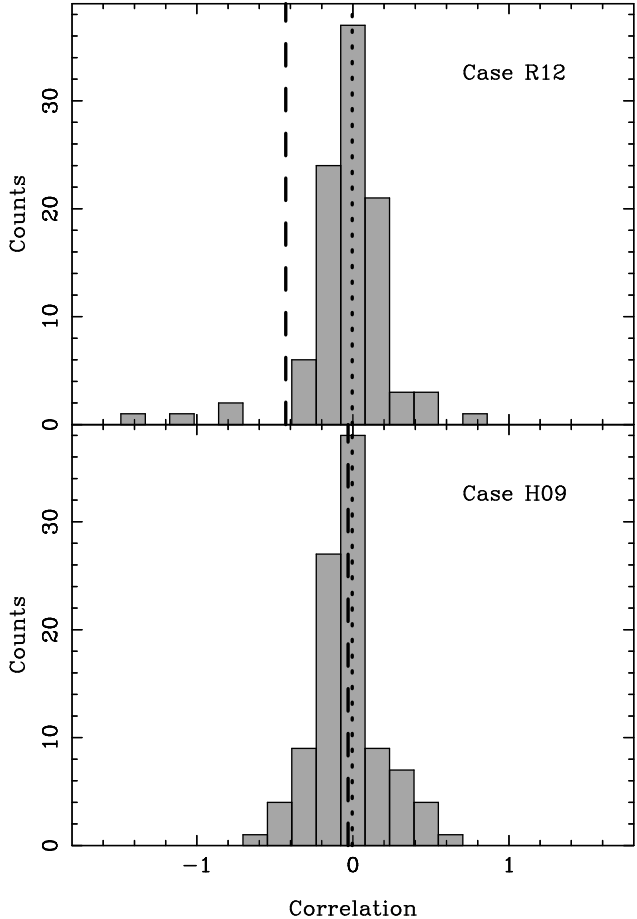}
\caption{Histograms showing the distributions of measurements of the correlation estimator $\tilde{\rho}_{pq}$ for 100 simulated ToA datasets for 
PSR J0437$-$4715 and PSR J0613$-$0200, in Case R12 (\textit{top}) and Case H09 (\textit{bottom}). See the text for details of the simulations. The 
Case R12 realization that included an extremely strong single GW source, as discussed in the text, resulted in a measurement of  $\tilde{\rho}_{pq}=-39.74$; 
this measurement is not shown in the Case R12 histogram. The vertical dashed line in each panel indicates the mean values of all 100 
estimated correlations in each case, and the vertical dotted line indicates the expected value of the correlation, $\rho_{pq}$, for an angular separation 
of $\theta_{pq}=49.8^{\circ}$.}
\end{figure}

The increased scatter in the Case R12 correlations with respect to the Case H09 correlations in Figure~7 is caused by outlying estimates in 
only a few realizations of ToAs. This is shown in Figure~8, where we display the histograms of the $\tilde{\rho}_{pq}$ measurements between simulated ToA datasets 
for PSR J0437$-$4715 and PSR J0613$-$0200 in each case. Correlation estimates $|\tilde{\rho}_{pq}|>1$ were possible because we normalized the estimated 
covariances between $D^{p}_{i}$ time-series using the expected cross-PSD between the time-series. While most measurements in both cases are concentrated 
about the expected value of $\rho_{pq}$, a few Case R12 measurements are significantly displaced. This is consistent with the results of \S4. 
We also stress that the large scatter of the estimator common to both cases is expected, and intrinsic to the GW signal.

\section{Discussion}

We have shown that the ToA variations induced by GWs from the predicted binary SMBH population are not consistent with the model described in \S2. 
The random Gaussian model for $\delta^{p}_{i}$ described in \S2 and approximated in Case H09 is reasonable given that a large number of GW sources are 
expected to contribute to the GW-induced ToA variations. That is, the values of $\delta^{p}_{i}$ at all times $t_{i}$ are the sums of many random variables. 
An argument based on the classical central limit theorem would suggest that $\delta^{p}_{i}$ would then be Gaussian random at every time $t_{i}$. 
It is apparent, however, that such a central limit theorem-based argument does not apply to the Case R12 realizations of $\delta^{p}_{i}$. 
This is because of the nature of the GW sources contributing to $\delta^{p}_{i}$ in Case R12. 

In Case R12, a few sources contribute most of the PSD of $\delta^{p}_{i}$ at every frequency, as shown in Figure~2. These sources are rare, because they are found at 
the high-$h_{0}$ tail of the $\Phi_{\text{fit}}$ source distribution. The estimators we consider in this work, $\tilde{S^{p}_{k}}$ and $\tilde{\rho}_{pq}$, are 
dominated in Case R12 by a few GW sources that need not occur in every realization of the $\delta^{p}_{i}$ time-series. This is why the distributions of these 
estimators are different between the cases. The quantities that we estimate, $S_{g}(f)$ and $\rho_{pq}$, are used to define the covariance matrix of the 
GW-induced ToA variations (see Equation~7). We have therefore shown that the ToA variations induced by GWs from binary SMBHs 
are dominated by the effects of a few strong, rare sources and cannot be accurately modeled using the random Gaussian process discussed in \S2.

\subsection{Implications of our results for experiments focused on a GW background}

Current PTA data analysis techniques use assumptions about the statistics of $\delta^{p}_{i}$ to attempt to estimate or constrain the amplitude of the 
characteristic strain spectrum of GWs from binary SMBHs. We consider the implications of our results for a selection of techniques in turn. We assume, in this 
discussion, that our results for the statistics of GW-induced ToA variations would apply even if the normalization of the GW characteristic strain spectrum 
$h_{\text{c},\,\text{fit}}(f)$, which we refer to as the GW amplitude,\footnote{We make a distinction between this amplitude and the $A_{1\,\text{yr}}$ parameter 
introduced in Equation~1, because $h_{\text{c},\,\text{fit}}(f)$ does not have exactly the same form as given in Equation~1.} were scaled up or down. 
Such a scaling could occur, for example, under different scenarios for whether coalescing SMBHs accrete gas before or after 
coalescence \citep{svc08}. 

We summarize a few key techniques here:
\begin{itemize}

\item \citet{jhl+05} describe a statistic which measures the degree of correlation between estimates of $\rho_{pq}$ from ToA data, and the expected 
functional form of $\rho_{pq}$. The expected detection significance, which is estimated under the assumption that the GW-induced ToA variations are 
Gaussian random, saturates at high values of the GW amplitude once the variance of the statistic is dominated by the stochasticity of the GW signal 
(see our Figure~8 and related discussion). 

\item \citet{jhv+06} constrain the amplitude of the GW characteristic strain spectrum from binary SMBHs by estimating the maximum possible GW signal present in 
measured data, under the assumption that the data could be modeled using a white noise process and GW-induced ToA variations. A statistic that 
estimates the GW background amplitude from individual pulsars was measured, and compared to the simulated distributions of the statistic for 
different GW amplitudes. The simulated statistic distributions were created from simulated ToA datasets with GW-induced ToA variations 
included using the {\sc tempo2} plugin GWbkgrd. 

\item \citet{vlm+09} present a Bayesian parameter estimation method for the GW characteristic strain spectral amplitude.\footnote{Though their method also estimates the 
spectral index of the GW characteristic strain spectrum, we assume marginalization over this parameter in our discussion here.} \citet{vlj+11} used 
this method to constrain the GW amplitude. This method requires an evaluation of the likelihood of the parameters used to model the ToA datasets, 
which include the GW amplitude. The likelihood is the probability distribution of the data given the model parameters.
The GW amplitude is used to calculate the covariance matrix, $\mathbf{C_{pq}}$, 
between the GW-induced ToA variations for pulsars $p$ and $q$ (see Equations 7, 8). This covariance matrix in turn is used to define the 
PTA likelihood, assuming that the GW-induced ToA variations can be modeled as a random Gaussian process.

\item \citet{dfg+12} use a PTA likelihood similar to the work of \citet{vlm+09} to constrain the GW amplitude by evaluating the distribution of a 
maximum likelihood estimator for the amplitude. They also use a method similar in concept to \citet{jhl+05} to attempt to detect the GW signal from binary SMBHs.

\end{itemize}

First, the non-Gaussianity of the GW-induced ToA variations means that 
the estimate of the intrinsic GW-induced variance of the \citet{jhl+05} statistic will be incorrect. This will affect estimates of the detection significance, particularly 
in the ``strong signal regime'', where the effects of GWs in the ToAs are large compared to all other noise processes. Second, the limit on the GW amplitude 
placed by \citet{jhv+06} will be biased. The \citet{jhv+06} limit was placed by finding the GW amplitude for which 95\% of simulated statistic values were 
above the measured value. The distribution of their statistic derived using our simulations would be different. 
Ruling out a GW amplitude using the \citet{jhv+06} technique does not necessarily rule out a GW signal corresponding to our simulations with the same 
confidence. Finally, our results indicate that the likelihoods evaluated by \citet{vlm+09} and \citet{dfg+12} will also be biased, leading to a similar effect on 
GW amplitude constraints made using their methods. 
A definitive statement on the magnitude of the consequences of our simulations for current constraints on the GW amplitude cannot be made, however, 
without fully considering the various PTA data analysis methods.

\subsection{Single GW source detection prospects}

We have established that in every frequency bin of the 5-year datasets we consider, a few strong GW sources dominate the 
PSD of $\delta^{p}_{i}$. This means that we cannot consider the expected GW signal from binary SMBHs to form a 
background.\footnote{This result is analogous to the case of the extragalactic background light \citep{dpr+11}, where the 
summed electromagnetic radiation from AGN and star-forming galaxies is dominated by strong individual sources, behind which 
myriad further objects combine to form an apparently isotropic background too uniform to be resolved by current telescopes.}
We briefly consider the prospects for there being single sources of GWs that are detectable by PTAs. Various methods of 
detecting and characterising individual continuous sources of GWs with PTAs have recently been presented 
\citep{yhj+10,bp10,cc10,lwk+11,bs12,esc12}. There are, however, few predictions for the expected numbers of detectable sources. 
\citet{svv09} analysed similar binary SMBH population models to those considered here to suggest that a 5-year ToA dataset would include 5$-$10 
single GW sources above the mean ``stochastic background'' level, mainly at GW frequencies greater than $10^{-8}$\,Hz. Their definition of a 
\textit{resolvable} source as one which has a (mean) strain amplitude that is greater than the mean background level is conservative. 
This is because a PTA is capable of spatial, as well as frequency resolution. The 
background contribution per spatial resolution element of a PTA will be less than the all-sky background level, resulting in a higher source amplitude 
to background ratio for a bright source located in the resolution element.

\begin{figure}
\centering
\includegraphics[angle=-90, scale=0.75]{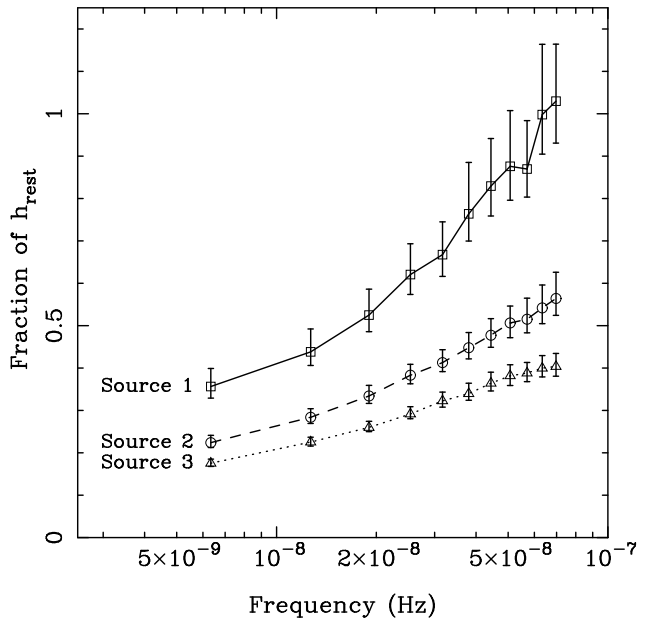}
\caption{The average strain amplitudes of the three highest-amplitude binary SMBHs in frequency bins with $0\leq k\leq10$ 
for each realization of the population. The strain amplitudes are expressed as fractions of the mean summed amplitude of the remaining 
sources. We also show the 5th and 95th percentiles of the strain amplitudes, with their deviations from the means scaled down by a 
factor of 10. We made 300 realisations of the source population to produce this figure. As indicated in the Figure, squares 
(the solid line) depict the mean amplitudes of the strongest sources, circles (the dashed line) depict the mean amplitudes of the second 
strongest sources, and triangles (the dotted line) depict the mean amplitudes of the third strongest sources.}
\end{figure}

The exact number of resolvable GW sources given a GW background level for PTAs depends on the particular search method. For example, 
\citet{bp10} suggest that a PTA composed of $N$ pulsars could resolve up to $2N/7$ sources per frequency bin. In Figure~9, we present a 
simple indication of the expected amplitudes of strong individual sources in the $0\leq k\leq11$ spectral bins of our fiducial 5-year dataset. 
Using 300 Case R12 realizations of the GW source population, we found the mean strain amplitudes in each spectral bin of the strongest three 
GW sources. We express these amplitudes as multiples of the mean summed strain amplitude, $h_{\text{rest}}$, of the 
remaining sources. The errors in the $h_{\text{rest}}$ values were not included in the error bars as they were very small.

If we consider the sources besides the strongest three in a spectral bin to form a ``background'',\footnote{This is by no means a rigorous definition of 
a background relative to the number of sources. The exact definition is dependent on the single source search method and the characteristics of the 
PTA.} it is clear that, for spectral bins with $k\geq2$, three sources, on average, produce the same total strain amplitude as the remaining sources. Even for 
the $k=0$ spectral bin, three sources are expected to produce more than half the total strain amplitude of the remaining sources. Indeed, the 
strongest source in the $k=0$ spectral bin has an average strain amplitude that is $\sim0.35h_{\text{rest}}$, which implies that a PTA which can resolve out two-thirds 
of the sky will detect equal contributions from the source and from the background. 


Blind searches for single GW sources with PTAs are therefore important. PTA data analysis methods that attempt to detect an isotropic component 
will not optimally recover the entirety of the GW signal from binary SMBHs, and could perhaps miss a large component of the signal for some realizations of the 
GW source population. A careful consideration of the efficacy of GW background detection methods as 
compared to search methods for single sources, given the predicted source characteristics, is required.

\subsection{Limitations of our approach to modelling the GW signal from binary SMBHs}


A shortcoming of our approach towards modeling $\delta^{p}_{i}$, and indeed of all predictions for the GW signal from binary SMBHs to date, is the assumption 
of circular orbits for all binaries. Recent work \citep[e.g.,][]{s10,pbb+11,kjm11} suggests that binary SMBHs emitting GWs in the PTA frequency 
regime will have highly eccentric orbits. The candidate binary SMBH OJ\,287 \citep[e.g.,][]{vlt+11} is in fact modeled with an orbital eccentricity of 
$\sim$0.7. The GW waveform of an eccentric binary radiating in the PTA band spans many frequencies, and does not follow the 
frequency-time relation of Equation~15, or the dependence of the GW strain amplitude on frequency of Equation~14. Therefore, if most binary SMBHs 
radiating in the PTA band are eccentric, the predicted mean spectral slope of the characteristic strain spectrum (Equation~1) will change. We will also need to 
account for the binary eccentricity distribution in our predictions of the statistics of GW-induced ToA variations. Other effects that require further 
investigation are the effects of gas and stars on binaries. We do not include these effects in our model because the understanding of their 
combined contributions towards specifying the binary SMBH population is not sufficiently advanced.

\section{Conclusions}

We have used a sophisticated model for galaxy evolution \citep{gwb+11} to predict the distribution of binary SMBHs radiating GWs in the PTA frequency band. 
By drawing lists of GW sources from this distribution, we simulate the effects of GWs from binary SMBHs on 5-year pulsar ToA datasets. We compare 
these simulations (Case R12) with simulated pulsar datasets containing the effects of an equivalent-amplitude GW signal modeled as a random 
Gaussian process (Case H09). We estimate the PSDs of the simulated GW-induced ToA variation time-series, and the correlations between these 
time-series for different pulsars. We find that the distributions of the PSD estimators of the realizations of the GW-induced ToA variations are different 
between the cases in every frequency bin, although the mean estimated PSDs are the same in each case. While in Case R12 the estimated PSDs 
are concentrated at lower values than in Case H09, the Case R12 estimations extend to higher PSD values than the Case H09 estimations. 
We also find that the functional form of the Hellings \& Downs curve is recovered on average in both cases. The correlations between the GW-induced 
ToA variation time-series for different pulsars in Case R12 are, however, significantly more scattered 
about the expected values than in Case H09. We interpret our results in terms of the influence of strong individual GW sources on the ToAs in Case R12. 

We conclude the following:
\begin{enumerate}
\item The effects of GWs from binary SMBHs on pulsar ToAs cannot be accurately modeled using existing methods, i.e, as a random Gaussian process. 
This is because a few GW sources dominate the PSD of the GW-induced ToA variations at all frequencies, with reducing numbers of sources contributing 
equivalent PSD fractions in higher frequency bins.
\item Our results directly affect existing PTA data analysis methods aimed at detecting or estimating the parameters of the GW signal from binary SMBHs. 
The projected detection significance will be biased. 
\item The prospects for single GW source detection are strong. Individual sources could potentially be resolved in all GW-dominated frequency bins of a 5-year dataset.
\end{enumerate}
We emphasize that future searches for GW signals from binary SMBHs in pulsar datasets need to be sensitive to both individual sources as well 
as a GW background.

\acknowledgments

VR is a recipient of a John Stocker Postgraduate Scholarship from the Science and Industry Endowment Fund. JSBW acknowledges an 
Australian Research Council Laureate Fellowship. GH is the recipient of an Australian
Research Council QEII Fellowship (project \#DP0878388). The Millennium and Millennium-II Simulation databases used in this paper and the web application 
providing online access to them were constructed as part of the activities of the German Astrophysical Virtual Observatory. We acknowledge use of the mpfit IDL 
routines of C. B. Markwardt. We thank W. Coles for his extensive and insightful comments into draft versions of the manuscript. We also thank the 
anonymous referee for comments that were useful in improving the manuscript.

\end{document}